\documentclass{article}
\usepackage{bm,amsmath,graphicx}

\title{Optimization of the Asymptotic Property of Mutual Learning Involving an Integration Mechanism of Ensemble Learning}
\author{Kazuyuki Hara$^{1}$ \thanks{E-mail:hara@tokyo-tmct.ac.jp}, Takahiro Yamada$^{2}$ } 
\date{$^{1}$Tokyo Metropolitan College of Industrial Technology \\Higashi-oi 1-10-40, Shinagawa-ku, Tokyo 140-0011.\\ $^{2}$Toyohashi University of Technology
\\1-1, Hibarigaoka, Tempaku, Toyohashi, Aichi, 441-8580.}
\begin{document}
\maketitle

\begin{small}
{\bf Abstruct--}
We propose an optimization method of mutual learning which converges
into the identical state of optimum ensemble learning within the
framework of on-line learning, and have analyzed its asymptotic property
through the statistical mechanics method.The proposed model consists of
two learning steps: two students independently learn from a teacher, and
then the students learn from each other through the mutual learning. 
In mutual learning, students learn from each other and the
generalization error is improved even if the teacher has not taken part 
in the mutual learning. However, in the case of different initial
overlaps(direction cosine) between teacher and students, a student with
 a larger initial overlap tends to have a larger generalization error 
than that of before the mutual learning. To overcome this problem, our proposed 
optimization method of mutual learning optimizes the step sizes of two 
students to minimize the asymptotic property of the generalization
error. 
Consequently, the optimized mutual learning converges to a
generalization error identical to that of the optimal ensemble
learning. 
In addition, we show the relationship between the optimum step size of 
the mutual learning and the integration mechanism of the ensemble learning.

\end{small}

\section{Introduction} 
As a model involving the interaction between students, Kinzel proposed mutual learning within the framework of on-line learning\cite{Klein2004,Kinzel2000,Kinzel2003}. Kinzel's model employs two students, and a student learns with the other student acting as a teacher. The target of his model is to obtain the same networks through the learning. On the other hand, ensemble learning algorithms, such as bagging\cite{Breiman1996} and Ada-boost\cite{Freund1997}, try to improve upon the performance of a weak learning machine by using many weak learning machines; such learning algorithms have recently received considerable attention. We have noted, however, that the mechanism of integrating the outputs of many weak learners in ensemble learning is similar to that of obtaining the same networks through mutual learning. 

From the point of view of the learning problem, how the student
approaches the teacher is important. However,
Kinzel\cite{Klein2004,Kinzel2000,Kinzel2003} does not deal with the
teacher-student relation since a teacher is not employed in his
model. In contrast to Kinzel's model, we have proposed mutual learning
between two students who learn from a teacher in
advance\cite{Hara2007}. In our previous work\cite{Hara2007}, we showed
that the generalization error of the students becomes smaller through
the mutual learning even if the teacher does not take part in the mutual
learning. We also showed that a student with a larger initial
overlap(direction cosine) for mutual learning transiently passes through
a state of the optimum ensemble learning when the limit of the learning 
step size is zero. 

In this paper, we propose a new mutual learning algorithm that uses a different learning step size for each student. We analyze the asymptotic property of the proposed learning algorithm through the statistical mechanics method, and propose an optimization method for the learning step size. By using the optimum learning step size, we can obtain the optimum asymptotic property of the generalization error through mutual learning. The proposed method is an expansion of our previous work\cite{Hara2007}. 

In this paper, we assume that each teacher and student is a linear perceptron. An on-line learning\cite{Saad1998} scheme is employed. In the proposed method, two students individually learn from a teacher during initial learning, and then they learn from each other during mutual learning. Therefore, we assume the overlaps between teacher and students are not zero at the initial state of mutual learning. In the mutual learning, each student learns from the other as the teacher. Since a teacher is not used in the mutual learning, we refer to a latent teacher in this paper. 

In Section 2, we formulate latent teacher, student, and mutual learning algorithms. In Section 3, we derive differential equations of the order parameters that depict the dynamics of mutual learning. We employ different learning step sizes for each student. We then derive the generalization error by using the order parameters. In Section 4, we solve the differential equations with different learning step sizes, and then analyze the effect of the learning step size on the asymptotic property of the mutual learning. After that, we obtain the optimum ratio of the students' learning step sizes which realizes the minimum generalization error. Moreover, we discuss the relation between the learning step size of mutual learning and the integration mechanism of ensemble learning. 

\section{Formulation of mutual learning with a latent teacher} 
\begin{figure}[t] 
\begin{center} 
\includegraphics[width=4cm]{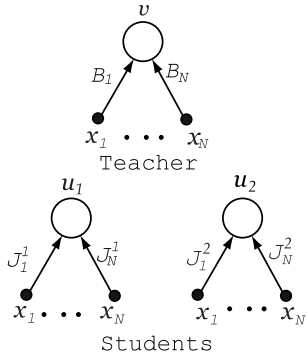} 
\end{center} 
\vspace{-.6cm} 
\caption{
\label{network}
Network structure of latent teacher and student networks, all having the same network structure.} 
\vspace{0.3cm} 
\end{figure}

In this section, we formulate the latent teacher and student networks, and the mutual learning algorithms. We assume the latent teacher and student networks receive $N$-dimensional input $\bm{x}(m) = (x_1(m) , \ldots , x_N(m))$ at the $m$-th learning iteration as shown in Fig. \ref{network}. Learning iteration $m$ is ignored in the figure. 
The latent teacher network is a linear perceptron, and the student networks are two linear perceptrons. We also assume that the elements $x_i(m)$ of the independently drawn input $\bm{x}(m)$ are uncorrelated random variables with zero mean and $1/N$ variance; that is, the elements are drawn from a probability distribution $P(\bm{x})$. In this paper, the thermodynamic limit of $N \rightarrow \infty$ is assumed. The size of input vector $|\bm{x}|$ then becomes one. 
\begin{equation} 
\langle x_i \rangle=0, \ \ \ \langle (x_i)^2 \rangle =\frac{1}{N}, \ \ \ |\bm{x}|=1, 
\label{input} 
\end{equation} 
\noindent 
where $ \langle \cdots \rangle$ denotes average, and $|\cdot|$ denotes the norm of a vector. 

The latent teacher network is a linear perceptron, and is not subject to training. Thus, the weight vector is fixed in the learning process. The output of the latent teacher $v(m)$ for $N$-dimensional input $\bm{x}(m)=(x_1(m), x_2(m), \ldots, x_N(m))$ at the $m$-th learning iteration is 
\begin{align} 
v(m)&= \sum_{i=1}^N B_i x_i(m) = \bm{B} \cdot 
\bm{x}(m), 
\label{teacher_output}\\ 
\bm{B}&=(B_1, B_2, \ldots, B_N), 
\end{align} 
\noindent 
where latent teacher weight vector $\bm{B}$ is an $N$-dimensional vector like the input vector, and each element $B_i$ of the latent teacher weight vector $\bm{B}$ is drawn from a probability distribution of zero mean and unit variance. Assuming the thermodynamic limit of $N \rightarrow \infty$, the size of latent teacher weight vector $|\bm{B}|$ becomes $\sqrt{N}$. 
\begin{equation} 
\langle B_i \rangle=0, \ \ \ \langle (B_i)^2 \rangle =1, \ \ \ |\bm{B}|=\sqrt{N}. 
\label{latent teacher_weight} 
\end{equation} 
\noindent 
The output distribution for the latent teacher $P(v)$ follows a Gaussian distribution of zero mean and unit variance in the thermodynamic limit of $N \rightarrow \infty$. 

The two linear perceptrons are used as student networks that compose the mutual learning machine. Each student network has the same architecture as the latent teacher network. Each element of $\bm{J}^k(0)$ which is the initial value of the $k$-th student weight vector $\bm{J}^k$ is drawn from a probability distribution of zero mean and unit variance. The norm of the initial student vector $|\bm{J}^k(0)|$ is $\sqrt{N}$ in the thermodynamic limit of $N \rightarrow \infty$, 
\begin{equation} 
\langle J^{k}_i(0) \rangle=0, \ \ \langle (J^{k}_i(0))^2 \rangle=1, \ \ |\bm{J}^k(0)|=\sqrt{N}. 
\label{student_weight} 
\end{equation} \noindent 
The $k$-th student output $u_k(m)$ for the $N$-dimensional input $\bm{x}(m)$ is 
\begin{align} 
u_k(m)&=\sum_{i=1}^N J^k_i(m) x_i(m) = \bm{J}^k(m) \cdot \bm{x}(m), \label{student_output}\\ \bm{J}^k(m)&=(J_1^k, J_2^k, \ldots, J_N^k). 
\end{align} 
\noindent 
Generally, the norm of student weight vector $|\bm{J}^k(m)|$ changes as the time step proceeds. Therefore, the ratio $l_k$ of the norm to $\sqrt{N}$ is considered and is called the length of student weight vector $\bm{J}^k$. The norm at the $m$-th iteration is $l_k(m) \sqrt{N}$, and the size of $l_k(m)$ is $O(1)$. 
\begin{equation} 
|\bm{J}^k(m)|=l_k(m) \sqrt{N} 
\label{length} 
\end{equation} 
\noindent 
The distribution of the output of the $k$-th student $P(u_k)$ follows a Gaussian distribution of zero mean and $l^2_k$ variance in the thermodynamic limit of $N \rightarrow \infty$. 

Next, we formulate the learning algorithm. After the students learn from a latent teacher, mutual learning is carried out. The learning equation of the mutual learning is 
\begin{equation} 
\bm{J}^{k}(m+1) = \bm{J}^{k}(m)+ \eta_k \Bigl(\ u_{k'}(m)-u_{k}(m) \ \Bigr) \bm{x}(m), 
\label{mutual_learning} 
\end{equation} 
\noindent 
where $k$ is 1 or 2 and $k \neq k'$. $m$ denotes the iteration number. Equation (\ref{mutual_learning}) shows that mutual learning is carried out between two students. Therefore, the teacher used in the initial learning is called a latent latent teacher. We use the gradient descent algorithm in this paper, while another algorithm was used in Kinzel's work \cite{Klein2004}. When the interaction between students is introduced, the performance of students may be improved if they exchange knowledge that each student has acquired from the latent teacher in the initial learning. In other words, two students approach each other through mutual learning, and tend to move towards the middle of the initial weight vectors. This tendency is similar to the integration mechanism of ensemble learning, so mutual learning may mimic this mechanism. 

\section{Theory} 
In this section, we first derive the differential equations of two order parameters which depict the behavior of mutual learning. After that, we derive an auxiliary order parameter which depicts the relationship between the latent teacher and students. We then rewrite the generalization error using these order parameters. 

We first derive the differential equation of the length of the student weight vector $l_k$. $l_k$ is the first order parameter of the system. We modify the length of the student weight vector in Eq. (\ref{length}) as $\bm{J}^k \cdot \bm{J}^k=Nl_k^2$ . To obtain a time dependent differential equation of $l_k$, we square both sides of Eq. (\ref{mutual_learning}). We then average the term of the equation using the distribution of $P(u_k,u_{k'})$. Note that $\bm{x}$ and $\bm{J}^k$ are random variables, so the equation becomes a random recurrence formula. We formulate the size of the weight vectors to be $O(N)$, and the size of input $\bm{x}$ is $O(1)$, so the length of the student weight vector has a self-averaging property. Here, we rewrite $m$ as $m=Nt$, and represent the learning process using continuous time $t$ in the thermodynamic limit of $N \rightarrow \infty$. We then obtain the deterministic differential equation of $l_k$, 
\begin{equation} 
\frac{d l_k^2}{dt}= (\eta_k^2-2\eta_k) l_k^2 + \eta_k^2 l_{k'}^2 - 2(\eta_k^2-\eta_k)Q. 
\label{differential_equation_l} 
\end{equation} 
\noindent 
Here, $k$ is 1 or 2, and $k \neq k'$. In this equation, $Q=q l_kl_{k'}$ and $q$ is the overlap between $\bm{J}^k$ and $\bm{J}^{k'}$, defined as 
\begin{equation} 
q=\frac{\bm{J}_k \cdot \bm{J}_{k'}}{|\bm{J}^k| \ |\bm{J}^{k'}|} = \frac{\bm{J}^k \cdot \bm{J}^{k'}}{Nl_kl_{k'}}, 
\label{definition_q} 
\end{equation} 
\noindent 
and $q$ is the second order parameter of the system. The overlap $q$ also has a self-averaging property, so we can derive the differential equation in the thermodynamic limit of $N \rightarrow \infty$. The differential equation is derived by calculating the product of the learning equation (Eq. (\ref{mutual_learning})) for $\bm{J}^k$ and $\bm{J}^{k'}$, and we then average the term of the equation using the distribution of $P(u_k,u_{k'})$. After that, we obtain the deterministic differential equation as 
\begin{equation} 
\frac{dQ}{dt}=(\eta_2-\eta_1\eta_2)l_1^2+(\eta_1 - \eta_1\eta_2)l_2^2 -(\eta_1+\eta_2-2\eta_1\eta_2)Q. 
\label{differential_equation_q} 
\end{equation} 
\noindent 
Equations (\ref{differential_equation_l}) and (\ref{differential_equation_q}) form closed differential equations. 

The analytical solutions of the length of the student $l_k$ and the overlap between students $Q$ are given by 
\begin{align} 
l_k^2(t)&=-A_1\frac{\eta_k}{\eta_{k'}}\exp(-(\eta_1+\eta_2)(2-(\eta_1+\eta_2))t)+(-1)^k 2A_2\frac{\eta_k}{\eta_2-\eta_1}\exp(-(\eta_1+\eta_2)t)+A_3, \label{l_ana} \\ 
Q(t)&=A_1 \exp(-(\eta_1+\eta_2)(2-(\eta_1+\eta_2))t)+A_2 \exp(-(\eta_1+\eta_2)t)+A_3, \label{Q_ana} 
\end{align} 
\noindent 
where 
\begin{align} 
A_1&=-\frac{\eta_1\eta_2(l_1^2(0)+l_2^2(0)-2Q(0))}{(\eta_1+\eta_2)^2},\\ 
A_2&=-\frac{(\eta_2-\eta_1)(\eta_2 l_1^2(0)-\eta_1 l_2^2(0)-(\eta_2-\eta_1)Q(0))}{(\eta_1+\eta_2)^2},\\
\ A_3&=\frac{\eta_2^2 l_1^2(0)+\eta_1^2 l_2^2(0)+2\eta_1 \eta_2 Q(0)}{(\eta_1+\eta_2)^2}. 
\end{align} 
\noindent 
$l_1(0)$ is the initial condition of student 1, and $l_2(0)$ is that of
student 2. $Q(0)=q(0)l(0)$, and $q(0)$ is the initial condition of the
overlap between student 1 and student 2. From Eqs. (\ref{l_ana}) and
(\ref{Q_ana}), $l_k^2(t)$ and $Q(t)$ converge to finite values at $t\rightarrow \infty$ if $2-(\eta_1+\eta_2)>0$ is satisfied. Then the convergence condition of $l_k^2(t)$ and $Q(t)$ is 
\begin{equation} 
\eta_1+\eta_2\ge 2. 
\end{equation} 

To depict the behavior of mutual learning with a latent latent teacher, we have to obtain the differential equation of overlap $R_k$, which is a direction cosine between latent teacher weight vector $\bm{B}$ and the $k$-th student weight vector $\bm{J}^k$ defined by Eq. (\ref{R}). We introduce $R_k$ as the third order parameter of the system. 
\begin{equation} 
R_k = \frac{\bm{B} \cdot \bm{J}^k}{|\bm{B}| \ |\bm{J}^k|} = \frac{\bm{B}\cdot \bm{J}^k}{Nl_k} 
\label{R} 
\end{equation} 
\noindent 
For the sake of convenience, we write the overlap between the latent teacher weight vector and the student weight vector as $r_k$ and $r_k=R_kl_k$. The differential equation of overlap $r_k$ is derived by calculating the product of $\bm{B}$ and Eq. (\ref{mutual_learning}), and we then average the term of the equation using the distribution of $P(v, u_k,u_{k'})$. The overlap $r_k$ also has a self-averaging property, and in the thermodynamic limit the deterministic differential equation of $r_k$ is then obtained through a calculation similar to that used for $l_k$. 
\begin{equation} 
\frac{dr_k}{dt}=\eta_k(r_{k'}-r_k) 
\label{differential_equation_r} 
\end{equation} 
The solution for overlap $r_k$ is obtained by solving simultaneous differential equations of Eq. (\ref{differential_equation_r}) for $k=1$ and $k'=2$, and for $k=2$ and $k'=1$. 
\begin{equation} 
r_k(t)=\frac{\eta_k(r_k(0)-r_{k'}(0))}{\eta_1+\eta_2}\exp(-(\eta_1+\eta_2)t) + \frac{\eta_2 r_1(0)+ \eta_1 r_2(0)}{\eta_1+\eta_2}, 
\label{r_ana} 
\end{equation} 
\noindent 
where $r_k(0)=R_k(0)l(0)$, and $R_k(0)$ is the initial overlap between the latent teacher and the $k$-th student. 

The squared error for the $k$-th student $\epsilon^k$ is then defined using the output of the latent teacher and that of the student as given in Eqs. (\ref{teacher_output}) and (\ref{student_output}), respectively. 
\begin{equation} 
\epsilon^k=\frac{1}{2}\Bigl(\bm{B}\cdot \bm{x}-\bm{J}^k \cdot \bm{x}\Bigr)^2 
\label{squared_error} 
\end{equation} 
\noindent 
The generalization error for the $k$-th student $\epsilon^k_g$ is given by the squared error $\epsilon^k$ in Eq. (\ref{squared_error}) averaged over the possible input $\bm{x}$ drawn from a Gaussian distribution $P(\bm{x})$ of zero mean and $1/N$ variance. 
\begin{align} 
\epsilon^k_g &= \int d\bm{x} P(\bm{x}) \ \epsilon^k\\
&=\frac{1}{2} \int d\bm{x}P(\bm{x})\Bigl(\bm{B}\cdot \bm{x}-\bm{J}^k \cdot \bm{x}\Bigr)^2. 
\label{generalization_error_in_input} 
\end{align} 
\noindent 
This calculation is the $N$-th Gaussian integral with $\bm{x}$ and it is hard to calculate. To overcome this difficulty, we employ coordinate transformation from $\bm{x}$ to $v$ and $u_k$ in Eqs. (\ref{teacher_output}) and (\ref{student_output}). Note that the distribution of the output of the students $P(u_k)$ follows a Gaussian distribution of zero mean and $l^2_k$ variance in the thermodynamic limit of $N \rightarrow \infty$. For the same reason, the output distribution for the latent teacher $P(v)$ follows a Gaussian distribution of zero mean and unit variance in the thermodynamic limit. Thus, the distribution $P(v,u_k)$ of latent teacher output $v$ and the $k$-th student output $u_k$ is 
\begin{align} 
P(v,u_k)&=\frac{1}{2 \pi \sqrt{|\Sigma|}} \exp \left [-\frac{(v, u_k)^T 
\Sigma^{-1} \ (v,u_k)}{2} \right ] \\ \Sigma&= \left ( 
\begin{array}{cc} 1 & r_k \\ r_k & l_k^2 
\end{array} \right) 
\end{align} 
\noindent 
Here, $T$ denotes the transpose of a vector, $r_k$ denotes $r_k=R_kl_k$, and $R_k$ is the overlap between the latent teacher weight vector $\bm{B}$ and the student weight vector $\bm{J}^k$ defined by Eq. (\ref{R}). Hence, by using this coordinate transformation, the generalization error in Eq. (\ref{generalization_error_in_input}) can be rewritten as 
\begin{align} 
\epsilon^k_g &= \frac{1}{2} \int dv du_k (v-u_k)^2 \\ 
&=\frac{1}{2}(1-2r_k+l_k^2). 
\label{eg} 
\end{align} 
\noindent 
Consequently, we calculate the dynamics of the generalization error by substituting the time step value of $l_k(t)$, $Q(t)$, and $r_k(t)$ into Eq. (\ref{eg}). 
\begin{align} 
\epsilon^k_g &= \frac{1}{2}\Biggl \{ 1-2\frac{\eta_k(r_k(0)-r_{k'}(0))}{\eta_1+\eta_2}\exp(-(\eta_1+\eta_2)t)- 2\frac{\eta_2 r_1(0)+ \eta_1 r_2(0)}{(\eta_1+\eta_2)} \Biggr. \nonumber \\ 
&+ \frac{\eta_k^2(l_1^2(0)+l_2^2(0)-2Q(0))}{(\eta_1+\eta_2)^2}\exp(-(\eta_1+\eta_2)(2-(\eta_1+\eta_2))t)\nonumber \\ 
&\Biggl. +(-1)^k \frac{2\eta_k(\eta_2l_1(0)-\eta_1l_2(0)-(\eta_2-\eta_1)Q(0))}{(\eta_1+\eta_2)^2}\exp(-(\eta_1+\eta_2)t) + \frac{\eta_2^2l_1^2(0)+\eta_1^2l_2^2(0)+2\eta_1\eta_2Q(0)}{(\eta_1+\eta_2)^2}\Biggr\} 
\label{eg_ana} 
\end{align} 
\section{Results} 
When the step sizes of two students are the same, the mutual learning asymptotically converges to the average weight vector of two students \cite{Hara2007}. In this section, we analyze the asymptotic property of mutual learning in the case of different step sizes, and then discuss the relationship between mutual learning and ensemble learning. 
\subsection{Effect of step size on the asymptotic property of mutual learning} 
\label{asymptotic_property} 
We analyze the effect of the learning step size on the asymptotic property of mutual learning. Two students use different learning step sizes. For this purpose, we use computer simulations. 

\begin{figure}[h] 
\begin{minipage}[t]{7.5cm} 
\begin{center} 
\includegraphics[width=7.5cm]{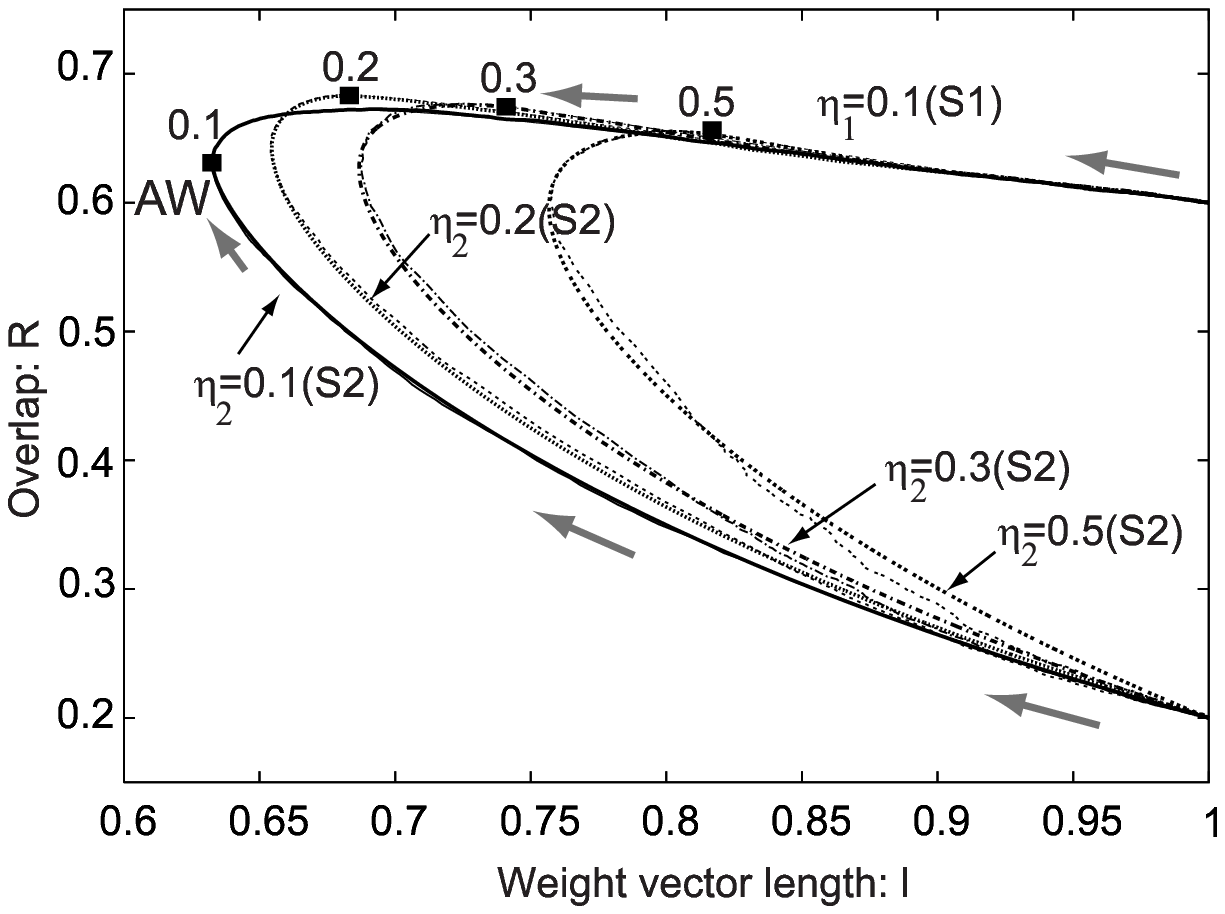} (a) $\eta_1=0.1$ 
\end{center} 
\end{minipage} 
\begin{minipage}[t]{7.5cm} 
\begin{center} 
\includegraphics[width=7.5cm]{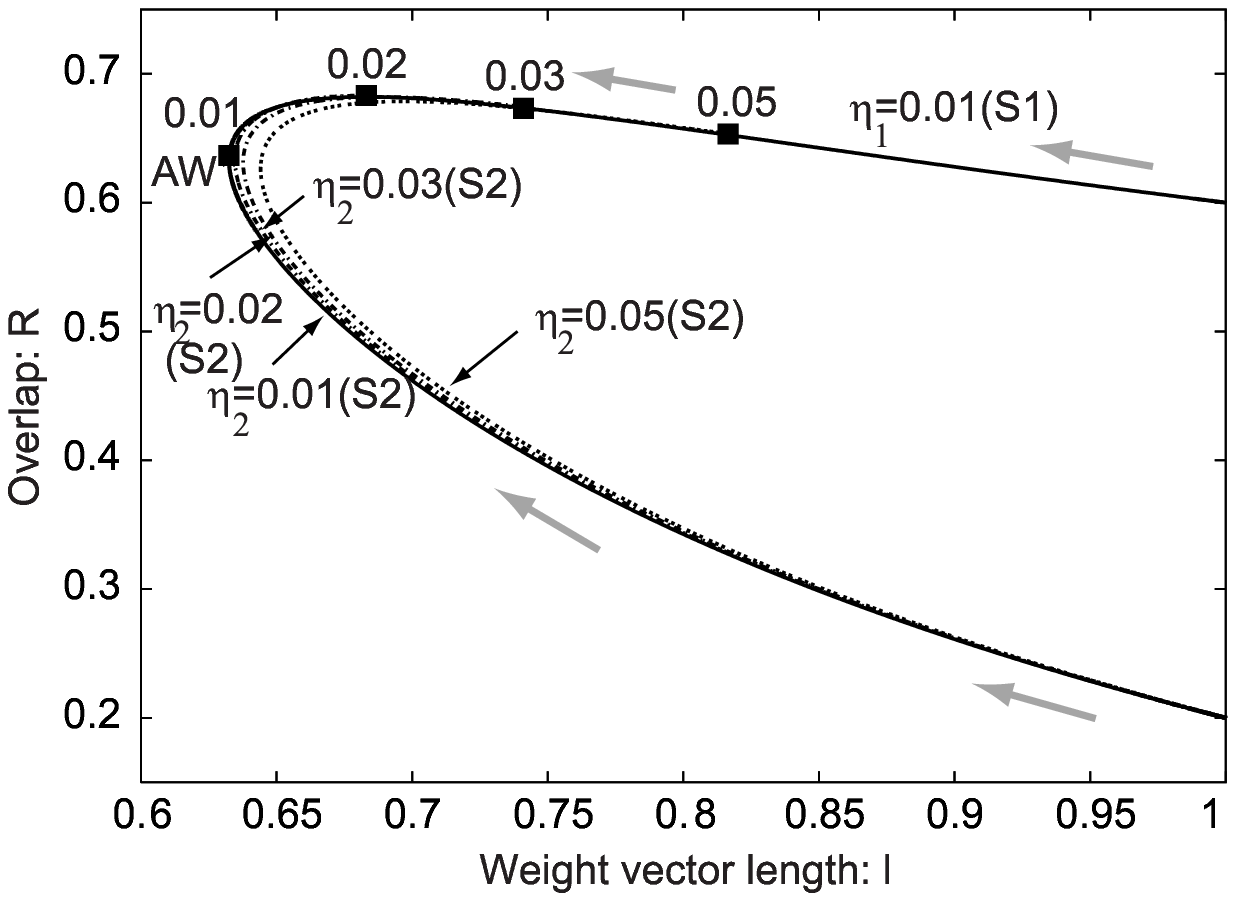} (b) $\eta_1=0.01$ 
\end{center} 
\end{minipage} 
\caption{
\label{ml_Rl}
Trajectories of student weight vector for the inhomogeneous case. The initial conditions were $l(0)=1$, $R_1(0)=0.6$, $R_2(0)=0.2$, and $q(0)=-0.2$. (a) Results of setting the learning step size to $\eta_1=0.1$(fixed) and $\eta_2=0.1, 0.2, 0.3$, or $0.5$. (b) Results of setting the learning step size to $\eta_1=0.01$(fixed) and $\eta_2=0.01, 0.02, 0.03$, or $0.05$.} 
\end{figure}

Figure \ref{ml_Rl} shows trajectories of the student weight vectors when
the initial overlaps between the latent teacher and the students were
inhomogeneous: (a) shows the results obtained through setting the
learning step size of student 1 ($\eta_1$) to 0.1(fixed), and setting
the learning step size of student 2 ($\eta_2$) to 0.1, 0.2, 0.3, or 0.5;
(b) shows the results obtained through setting the learning step size
$\eta_1$ to 0.01(fixed), and setting $\eta_2$ to 0.01, 0.02, 0.03, or
0.05. In these figures, the horizontal axis shows the length of the
student weight vector $l_k$, and the vertical axis shows the overlap
$R_k$. The initial conditions were $l_1(0)=l_2(0)=1$, $R_1(0)=0.6$,
$R_2(0)=0.2$, and $q(0)=-0.2$. The theoretical results obtained using
Eqs. (\ref{l_ana}), (\ref{Q_ana}), and (\ref{r_ana}) are shown as thick
lines, and the results obtained through computer simulations for
$N=10000$ are shown as thin lines. The upper lines show trajectories of
the weight vector of student 1, and the lower lines show trajectories of
the weight vector of student 2. The symbols of black rectangles show 
convergence points of trajectories of the student weight vectors. The numbers above the symbols show the learning step sizes of student 2. 

When the learning step sizes $\eta_1$ and $\eta_2$ were the same, student 1 started at $l_1(0)=1$ and $R_1(0)=0.6$, and converged to the average weight vector of the initial student vectors denoted by $\bm{AW}$. Student 2 started at $l_2(0)=1$ and $R_2(0)=0.2$, and also converged to the average weight vector denoted by $\bm{AW}$ when using the same learning step sizes. 

When the learning step sizes $\eta_1$ and $\eta_2$ were not the same, the convergence points were changed by using a different step size $\eta_2$ of $0.2, 0.3$, or $0.5$ as shown in Fig. \ref{ml_Rl}(a). As in Fig. \ref{ml_Rl}(a), Fig. \ref{ml_Rl}(b) shows that the convergence points were changed by using a different step size $\eta_2$ of $0.02, 0.03$, or $0.05$. Note that the convergence points for the same ratio of the learning step size tend to be the same. Thus, we pay attention to the effect of the ratio of learning step sizes $\eta_2/\eta_1$ in the mutual learning. 

\begin{figure}[h] 
\begin{center} 
\includegraphics[width=7.5cm]{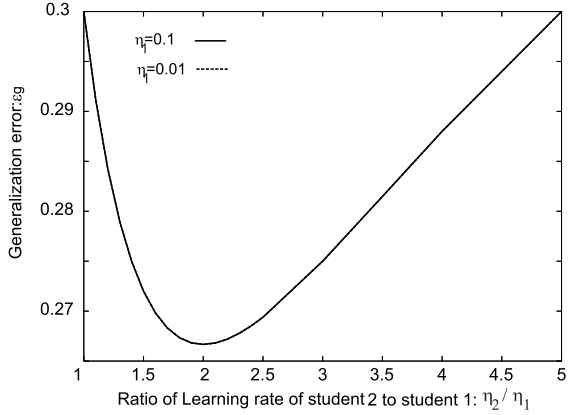} 
\end{center} 
\caption{
\label{ml_eg}
Relation between learning step size and generalization error. The learning step size of student 1 was 0.1 or 0.01(fixed), and that of student 2 was changed. The generalization error is minimized when the ratio of the learning step size is two for both cases. The optimum ratio is independent of the size of the learning step size.} 
\end{figure}

Figure \ref{ml_eg} shows the learning step size dependence of the generalization error. The learning step size of student 1 was 0.1 or 0.01(fixed), and that of student 2 was changed as shown in the figure. The horizontal axis shows the ratio of learning step sizes $\eta_2/\eta_1$, and the vertical axis shows the asymptotic property of the generalization error $\epsilon_g$. The asymptotic property of the generalization error is obtained using Eq. (\ref{eg_ana}) for the case of $t\rightarrow \infty$. The results show that the asymptotic property of the generalization error was minimized when the ratio $\eta_2/\eta_1$ was $2$. Consequently, the asymptotic property of the generalization error can be minimized by using the optimal ratio of learning step sizes. Next, we will obtain this optimal ratio of learning step sizes that minimizes the asymptotic property of the generalization error. 

\subsection{Optimization of the asymptotic property of the generalization error} 
We now analyze the asymptotic property of the generalization error based on the ratio of learning step sizes, and then we obtain the optimum ratio of learning step sizes $\eta_2/\eta_1$ that minimizes the asymptotic property of the generalization error. 

The asymptotic property of the order parameters is obtained by substituting $t \rightarrow \infty$ into Eqs. (\ref{l_ana}), (\ref{Q_ana}), and (\ref{r_ana}): 
\begin{align} 
l_1^2(\infty)&=l_2^2(\infty)=Q(\infty)=\frac{\eta_2^2l_1^2(0)+\eta_1^2l_2(0)+2 \eta_1\eta_2Q(0)}{(\eta_1+\eta_2)^2}, 
\label{l_Q_infty}\\ r_1(\infty)&=r_2(\infty)=\frac{\eta_2}{\eta_1+\eta_2}r_1(0)+\frac{\eta_1}{\eta_1+\eta_2}r_2(0). 
\label{r_infty} 
\end{align} 
\noindent 
The above equations show that the mutual learning converges to the internal dividing point of the initial student weight vectors. Using Eqs. (\ref{l_Q_infty}) and (\ref{r_infty}), we can obtain the asymptotic property of the generalization error: 
\begin{equation} 
\epsilon_g(\infty)=\frac{1}{2}\left\{ 1-2 \frac{\eta_2 r_1(0)+\eta_1 r_2(0)}{\eta_1+\eta_2} + \frac{\eta_2^2l_1^2(0)+\eta_1^2l_2(0)+2 \eta_1\eta_2Q(0)}{(\eta_1+\eta_2)^2}\right\}
\label{eg_infty} 
\end{equation} 
\noindent 
We rewrite the generalization error by replacing the ratio $\eta_2/\eta_1$ with $\alpha$: 
\begin{equation} 
\epsilon_g(\infty)=\frac{1}{2}\Biggl\{ 1-2\frac{\alpha r_1(0)+r_2(0)}{\alpha+1}+\frac{\alpha^2l_1^2(0)+l_2^2(0)+2\alpha Q(0)}{(\alpha+1)^2}\Biggr\}. 
\label{asymptotic_eg_ml} 
\end{equation} 
\noindent 
When the generalization error is minimized, $\partial \epsilon_g(\infty)/\partial \alpha=0$ is satisfied, so 
\begin{equation} 
\frac{\partial \epsilon_g}{\partial \alpha}=\frac{2\alpha l_1^2(0)+2Q(0)}{(\alpha+1)^2}-\frac{2(\alpha^2 l_1^2(0)+l_2^2(0)+2\alpha Q(0))}{(\alpha+1)^2}+\frac{2(\alpha r_1(0)+r_2(0))}{(\alpha+1)^2}-\frac{2r_1(0)}{\alpha+1}=0 
\label{condition_of_alpha} 
\end{equation} 
\noindent 
Solving Eq. (\ref{condition_of_alpha}), we obtain $\alpha^{opt}$ as 
\begin{equation} 
\alpha^{opt}=\frac{l_2^2(0)-Q(0)+r_1(0)-r_2(0)}{l_1^2(0)-Q(0)-r_1(0)+r_2(0)}. 
\label{alpha} 
\end{equation} 
\noindent 
Therefore, the optimum ratio of the learning step size is obtained through Eq. (\ref{alpha}). The optimum asymptotic property of the generalization error is obtained by substituting Eq. (\ref{alpha}) into Eq. (\ref{asymptotic_eg_ml}): 
\begin{equation} 
\epsilon_g^{opt}(\infty)=\frac{1}{2}\left\{1-2 (\kappa r_1(0)+(1-\kappa) r_2(0)) + \kappa^2l_1^2(0)+(1-\kappa)^2l_2^2(0)+2\kappa(1-\kappa)Q(0)\right\}. 
\label{optimal_eg} 
\end{equation} 
\noindent 
Here, $\kappa$ is defined as $\kappa=\alpha^{opt}/(1+\alpha^{opt})$. 

On the other hand, we can consider the linear combination of the initial
weight vectors of the students --- that is, $\bm{J}=C\bm{J}^1(0) + (1-C)
\bm{J}^2(0)$ --- and minimize the generalization error by $C$. This is
an ensemble learning with two students, so from the appendix, the optimum $C^*$ that minimizes the generalization error is 
\begin{equation} 
C^*=\frac{l_2^2(0)-Q(0)+r_1(0)-r_2(0)}{l_1^2(0)+l_2^2(0)-2Q(0)}.
\label{optimum_C} 
\end{equation} 
\noindent 
Therefore, the optimum ratio $C^*/(1-C^*)$ is obtained as 
\begin{equation} 
\frac{C^*}{1-C^*}=\frac{l_2^2(0)-Q(0)+r_1(0)-r_2(0)}{l_1^2(0)-Q(0)-r_1(0)+r_2(0)}=\frac{\eta_2^{opt}}{\eta_1^{opt}}, 
\end{equation} 
\noindent 
and it is shown that the optimum ratio of the learning step size of
mutual learning $\alpha^{opt}=\eta_2^{opt}/\eta_1^{opt}$ is equal to
that of the optimum linear combination of the initial weight vectors
$C^*/(1-C^*)$. Consequently, mutual learning using an optimum ratio of
learning step sizes converges to the optimum ensemble learning that is
the linear combination of the initial student vectors. 

\section{Conclusion} 
We have proposed an optimization method for mutual learning by means of minimizing the asymptotic property of the generalization error within the framework of on-line learning. We first formulated mutual learning with a latent teacher, and then derived the differential equations of order parameters that depict the learning process. The order parameters of mutual learning are the length of the student weight vector $l_k$ and the overlap between students $q$. To depict the relationship between the latent teacher and the students, we introduced the order parameter $R_k$. We derived these differential equations using statistical mechanics methods and solved them analytically. After that, we obtained the dynamics of the generalization error using these order parameters. 

Next, we used the theoretical results to analyze the relationship between the asymptotic property of the mutual learning and the learning step size of the students. From the results, we found that the asymptotic property of the mutual learning related to the ratio of the learning step sizes of two students, and was not related to the learning step size itself. We obtained the optimum ratio of the learning step size which minimizes the generalization error analytically. We also showed that the optimum ratio of the learning step sizes of the mutual learning is equal to the inverse of the ratio of optimum weights for an average of the linear combination of initial student weight vectors. We conclude that the integration mechanism of ensemble learning can be mimicked through mutual learning by introducing the interaction between students. Our future work will include analysis of the mutual learning with non-linear perceptrons. 

\section*{Acknowledgment} 
We would like to thank Masato Okada (The University of Tokyo) and Seiji Miyoshi (Kobe City College of Technology) for their useful discussions. Part of this study has been supported by a Grant-in-Aid for Scientific Research (C) No. 16500146.

\appendix 
\section{Ensemble learning} 
\label{ensemble} 
Ensemble learning is a learning method using many weak learning machines
to improve upon the performance of a single weak learning
machine\cite{Breiman1996,Freund1997,lazarevic2002}. Students learn from
the teacher individually, and then an ensemble output is calculated by
integrating the students' outputs. Because many students are used,
ensemble learning is effective when the students differ from each
other. Therefore, we assume that the overlap(direction cosine) between the $k$th student and the $k'$th student $q_{kk'}$ is not one. The ensemble output of the student networks $\overline{u}$ is given by the weighted average of each student output using the weights for averaging $C_k$: 
\begin{align} 
\overline{u}&=\sum_{k=1}^K C_ku_k=\sum_{k=1}^K C_k\Bigl(\bm{J}^k \cdot \bm{x}\Bigr) 
\label{weight_CA} 
\end{align} 
\noindent 
Here, the number of students is $K$ and we assume $\sum_{k=1}^K C_k=1$. In the following, we assume that the number of students is two. We use linear perceptrons as the students, so the average output of the two students is equal to the output of a perceptron having the average of the two student weight vectors. The weighted average of the two student weight vectors $\bm{J}^E$ is defined as follows\cite{Hara2007}. 
\begin{equation} 
\bm{J}^E=C_k \bm{J}^k+C_{k'}\bm{J}^{k'}=C \bm{J}^k+(1-C) \bm{J}^{k'}
\label{weight_ensemble_A} 
\end{equation} 
\noindent 
Here, we rewrite $C_k$ as $C$ and $C_{k'}$ as $1-C$ from $C_k+C_{k'}=1$. From this equation, ensemble learning can be viewed as the linear combination of the two student weight vectors. Note that ensemble learning is a static process, so there is no dynamical property. The length of the weight vector $l^E$ and the overlap $r^E$ are given by 
\begin{align} 
(l^{E})^2&=C^2l_k^2+(1-C)^2l_{k'}^2 + 2C(1-C)Q 
\label{l_ensemble_A}\\ 
r^{E}&=C r_k+(1-C)r_{k'} 
\label{r_ensemble_A} 
\end{align} 
\noindent 
The generalization error of ensemble output $\epsilon_g^E$ is given by substituting Eqs. (\ref{l_ensemble_A}) and (\ref{r_ensemble_A}) into Eq. (\ref{eg}): 
\begin{align} 
\epsilon_g^E&=\frac{1}{2}\left(1-2 r^E + (l^E)^2 \right) \nonumber \\ 
&= \frac{1}{2} \Bigl\{ \ 1-2(C r_k+(1-C)r_{k'})+C^2l_k^2+(1-C)^2l_{k'}^2+2C(1-C)Q \Bigr \}. 
\label{eg_opt_A} 
\end{align} 
\noindent 
If the optimum weight for average $C^*$ satisfies the condition of $\partial \epsilon^*_g/\partial C^*=0$, we obtain 
\begin{equation} 
C^*=\frac{l_{k'}^2-Q+r_{k}-r_{k'}}{l_k^2+l_{k'}^2-2Q} 
\label{CK_A} 
\end{equation} 
\noindent 
When the student weight vector length $l_k=l_{k'}=l$ and the overlap between the students $r_k=r_{k'}=r$, from Eq. (\ref{CK_A}) we obtain $C^*=(1-C^*)=1/2$ and the simple average of the two students is the optimum ensemble output. 

\begin{thebibliography}{99} 
\bibitem{Breiman1996} L. Breiman, Bagging predictors, {\it Machine Learning}, vol. 24, pp. 123-140 (1996). 
\bibitem{Freund1997} Y. Freund and R. E. Shapire, J. Comput. Syst. Sci. {\bf 55} (1997) 119. 
\bibitem{Saad1998} On-line Learning in Neural Networks, ed. D. Saad (Cambridge University Press, Oxford, 1998). 
\bibitem{Sollich1997} A. Krogh and P. Sollich, Phys. Rev. E, {\bf 55} (1997) 811. 
\bibitem{Hara2004} K. Hara and M. Okada, Neural Networks, {\bf 17} (2004) 215. 
\bibitem{Hara2005} K. Hara and M. Okada, J. Phys. Soc. Jpn. {\bf 74} (2005) 2966. 
\bibitem{Miyoshi2005} S. Miyoshi, K. Hara, and M. Okada, Phys. Rev. E, {\bf 71} (2005) 036116. 
\bibitem{lazarevic2002} A. Lazarevic and Z. Obradivic, Distributed and parallel databases, vol.11, pp. 203 (2002). 
\bibitem{Klein2004} Klein, E., et. al., Proc. Neural Inf. Pro. Sys. (2004). 
\bibitem{Kinzel2000}R. Metzler, W. Kinzel, and I. Kanter: Phys. Rev. E 62 (2000) 2555. 
\bibitem{Kinzel2003} R. Mislovaty, E. Klein, I. Kanter, and W. Kinzel: Phys. Rev. Lett. 91 (2003) 118701. 
\bibitem{Hara2007} Hara K. and M. Okada, J. Phys. Soc. Jpn. {\bf 76} (2007) 014001. 
\end{thebibliography}
\end{document}